# On the dependency of the parameters of fatigue crack growth from the fractal dimension of rough crack profiles

M. Paggi[1] and O. Plekhov[2]

[1]*Politecnico di Torino, Department of Structural, Geotechnical and Building Engineering*
*Corso Duca degli Abruzzi 24, 10129 Torino, Italy*

[2]*Institute of Continuous Media Mechanics of Russian Academy of Sciences*
*Ak. Koroleva str 1, 614013 Perm, Russia*

**Abstract**

A theoretical study based on dimensional analysis and fractal geometry of crack profiles is proposed to establish the relation between their fractal dimension $D$ (1<$D$<2) and the parameters defining the fatigue crack propagation rate. The exponent $m$ of the Paris' law is found to be an increasing function of the fractal dimension of the crack profile, $m = 2D/(2-D)$. This trend is confirmed by a quantitative analysis of fractographic images of Titanium alloys with different grain sizes (different roughness of crack profiles), by a new experimental test and by an indirect estimation of $D$ from crack growth equations accounting from crack-size effects in Steel and Aluminum. The present study can be considered as the first quantitative analysis of fractographic images aiming at relating the morphological features of cracks to their kinetics in fatigue.

Corresponding Author: Marco Paggi, marco.paggi@polito.it, Tel: +39-011-0904910, Fax: +39-011-0904899

**Nomenclature**

| Symbol | Definition |
|---|---|
| $a$ | crack length |
| $a_0$ | initial defect size |
| $a_f$ | critical (final) crack length |
| $C$ | coefficient in the Paris' law |
| $D$ | fractal dimension |
| $E$ | Young's modulus |
| $G$ | scaling constant |
| $h$ | characteristic length |
| $\Delta K$ | stress-intensity factor range |
| $K_{IC}$ | fracture toughness |
| $K_{IC}^*$ | renormalized fracture toughness |
| $\Delta K^*$ | renormalized stress-intensity factor range |
| $m$ | exponent of the stress-intensity factor range in the Paris' law |
| $N$ | number of cycles to failure |
| $R$ | loading ratio |
| $S$ | structure function |
| $z$ | vertical coordinate of the crack profile |
| $\alpha_i$ | incomplete similarity exponents |
| $\gamma$ | exponent in the Frost-Dugdale law |
| $\delta$ | sampling interval |
| $\sigma_y^*$ | renormalized yield strength |
| $\Delta \sigma$ | stress range |
| $\sigma_y$ | yield strength |

## 1. Introduction

The problem of crack growth is one of the key problems in fracture mechanics. The process of crack propagation is accompanied by the collective phenomena in mesodefect ensemble and multiscale damage accumulation that lead to the formation of geometrical complex crack profiles. The investigation of those profiles originated by monotonic loading in different materials (metals, polymers, glasses) shows that they exhibit statistical self-similar properties [1-4]. The crack profiles can be described using either statistical methods [5-8], or fractal geometry [9-14]. In this context, the latter method is particularly useful, since it permits to describe such complex topologies with a limited number of parameters, among which the fractal dimension is the most important one. Preliminary experimental results in [3] show that the fractal dimension of fractal profiles of Aluminum alloys ranges from 1.2 to 1.5, that of ZnSe ranges from 1.1 to 1.4, that of 4340 Steel ranges from 1.3 to 1.5, that of 300 Grade Maraging Steel ranges from 1.1 to 1.4 and finally that of Ti ranges from 1.0 to 1.1.

In case of fatigue crack growth, the crack profiles are much smoother than for monotonic loading and the existence of multiscale (fractal) characteristics has to be ascertain by high resolution microscopy (to visualize a trace of an emerging crack on a plane) or profilometers (to quantify roughness of a crack surface). Hence, fractal properties of fatigue cracks have been assessed only recently, after the pioneering study by Williford [3].

This work is devoted to the investigation of the effect of scaling properties of fatigue cracks on the value of the Paris' law parameters. This is motivated by the idea that the initial material heterogeneity (for instance, the grain size) controls the ability of the material to resist crack initiation or propagation, determines the self-similar properties of crack profiles and, as a main outcome, affects the fatigue crack growth rate and the fatigue life of the component.

The reason for the application of fractal geometry to fatigue crack growth can be easily shown by the analysis of cracks in failed specimens due to fatigue presented in Fig. 1(a) and adapted from [15]. The original profiles have been modified by removing the long-wave oscillation of the crack path in order to isolate roughness. As it can be readily seen, the crack profiles of Titanium (Ti) specimens are largely dependent on the grain size. Qualitatively speaking, the finer the material microstructure, as in case of the ultrafine grain Ti, the smoother the crack path. To quantify this effect in terms of fractality, the fractal dimension can be computed using the structure function $S$ which is defined as follows [16, 17]:

$$S(\delta) \cong \lim_{N \to \infty} \frac{1}{N} \sum_{i=1}^{N} [z(x_i) - z(x_i + \delta)]^2 = G^{2(D-1)} \delta^{(4-2D)} \qquad (1)$$

where $\delta$ is the sampling interval and $z$ is the vertical coordinate of the profile sampled in correspondence of discrete abscissa $x_i$. For a fractal profile, the $S$ vs. $\delta$ data obey a power-law, where the exponent $(4-2D)$ is related to the fractal dimension $D$. The fractal dimension ranges between 1 and 2 since we have an invasive fractal or, in other words, a profile with dimension intermediate between those of a line and of a surface. The scaling constant $G$ is related to the r.m.s. values of the sampled heights and is the other scale-independent parameter of the rough profile.

Using Eq. (1) and the profile data in Fig. 1(a), the plots in Fig. 1(b) are determined. The structure functions, in this bilogarithmic diagram, are very close to straight lines. Their slopes can be used to compute the fractal dimensions $D$ using the expected scaling law in Eq. (1). From the analysis of these data we find $D=1.24$ for coarse grain Ti and $D=1.14$ for ultrafine grain Ti. It is interesting to note that a finer material microstructure allows us to reduce the roughness of the crack profile, as can also be visually observed by comparing the two curves in Fig. 1(a). More importantly, in terms of effect on fatigue, a finer material microstructure increases the fatigue life (number of cycles corresponding to the specimen failure) up to 8% as compared to the standard coarse grain Ti [15]. Therefore, the fractality of rough cracks is a fundamental quantity for the kinetics of the process of crack growth whose effect has to be carefully understood.

This paper is structured as follows: in Section 2, a dimensional analysis approach combined with fractality is proposed for the first time to derive fatigue laws explicitly incorporating the properties of the crack profile morphology on the fatigue crack growth. Then, in Section 3, a quantification of the effect of the fractality of cracks on the fatigue life of a tested component is proposed by integrating the obtained crack-size dependent Paris' law. Finally, an experimental assessment of the proposed model is proposed in Section 4, along with a critical comparison with the existing scaling laws available in the literature.

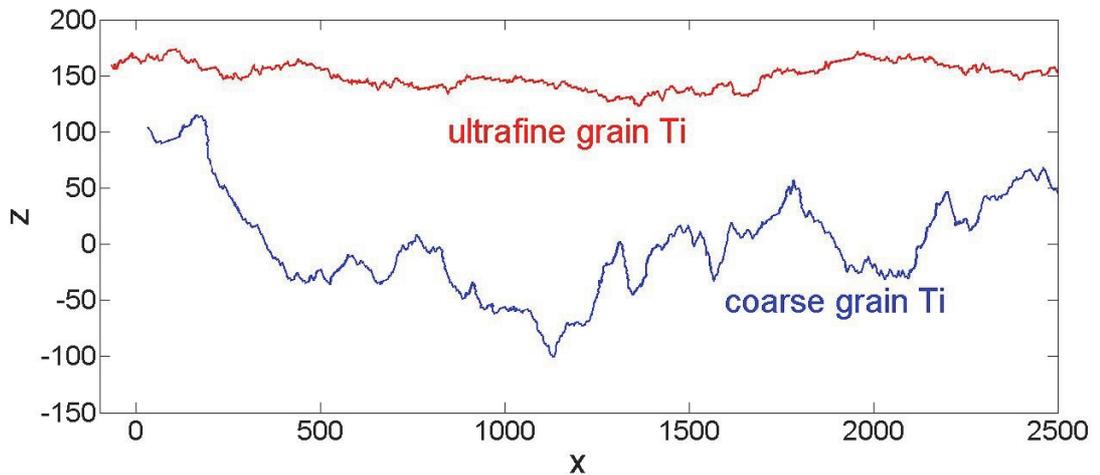

(a) Fatigue crack profiles of coarse grain and ultrafine grain Titanium (dimensions in micrometers).

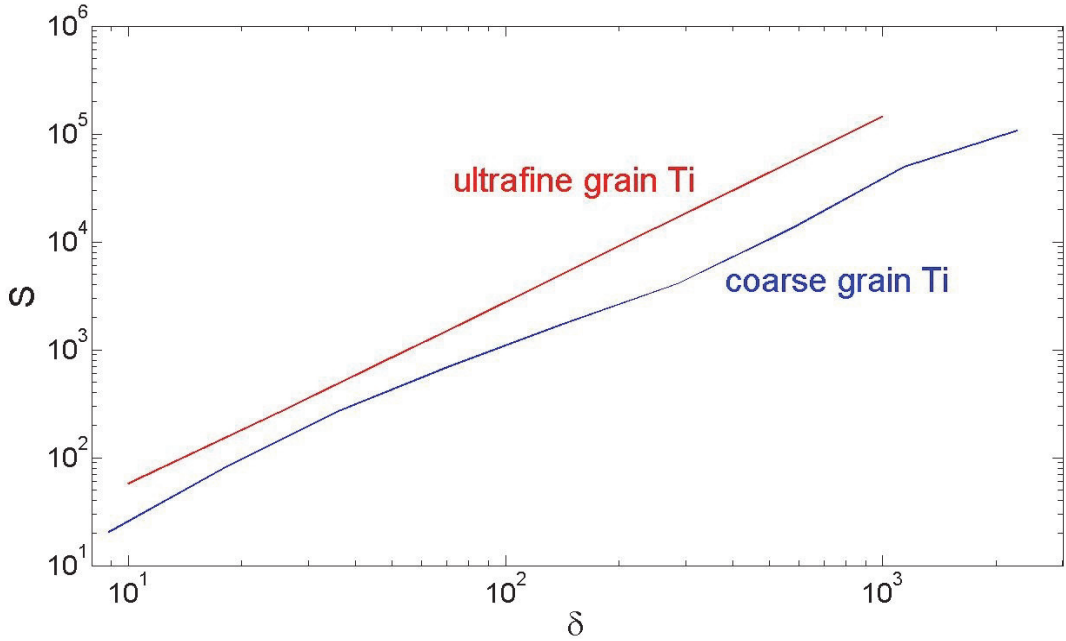

(b) Structure functions of the profiles in Fig. 1(a).

Fig. 1: (a) Characteristic crack profiles of coarse grain and ultrafine grain Titanium. Experimental data adapted from [12]. (b) Structure functions used to compute the fractal dimension $D$ of the two profiles.

## 2. Dimensional analysis considerations applied to fractal quantities

A dimensional analysis of the laws used to describe the Cumulative Fatigue Damage (CFD) and the Fatigue Crack Propagation (FCP) approaches has been proposed in [18,19]. Further applications have regarded the study of size-scale effects in fatigue [20], the grain-size effects [21], and finally the unification of the fatigue laws for quasi-brittle materials [22]. A general dependence on the most important variables affecting fatigue crack growth was considered in those studies. However, the classical Euclidean (nominal) quantities for the crack length and the stress-intensity factor were retained. In summary, the following functional dependence was put forward for the crack growth rate $da/dN$:

$$\frac{da}{dN} = F\left(\Delta K, E, K_{IC}, \sigma_y, a, 1-R\right) \qquad (2)$$

where $\Delta K$ is the stress-intensity factor range, $E$ is the Young's modulus, $K_{IC}$ is the fracture toughness, $\sigma_y$ is the yield strength, $a$ is the crack length and $R$ is the loading ratio. The application of the Buckingham's $\Pi$ Theorem [23] to reduce the number of variables involved in the problem leads:

$$\frac{da}{dN} = \left(\frac{\Delta K}{E}\right)^2 \Phi\left(\frac{K_{IC}}{\Delta K}, \frac{\sigma_y}{E}, \frac{E^2}{K_{IC}^2}a, 1-R\right) = \left(\frac{\Delta K}{E}\right)^2 \Phi\left(\Pi_1, \Pi_2, \Pi_3, \Pi_4\right) \qquad (3)$$

The assumption of incomplete similarity in the various dimensionless numbers allow us to obtain a crack size-dependent crack growth law (the reader is referred to [18-20] for more details about the validity of such a hypothesis):

$$\frac{da}{dN} = \left(\frac{\Delta K}{E}\right)^2 \left(\frac{K_{IC}}{\Delta K}\right)^{\alpha_1} \left(\frac{E^2}{K_{IC}^2}a\right)^{\alpha_2} (1-R)^{\alpha_3} \Phi_1 = \Delta K^{2-\alpha_1} a^{\alpha_2} (1-R)^{\alpha_3} \frac{\Phi_1}{K_{IC}^{-\alpha_1+2\alpha_2} E^{2(1-\alpha_2)}} \tag{4}$$

The main drawback of this approach consists in the fact that the incomplete similarity exponents $\alpha_i$ can only be determined according to experimental or numerical results and not from theoretical considerations. If, on the one hand, the classical Paris' law suggests $m = 2 - \alpha_1$, and several variants of that law support the power-law dependence on $(1-R)$, nothing can be said about $\alpha_2$. Actually, as we demonstrate in the sequel, it is possible to theoretically relate $\alpha_2$ to the fractal dimension of rough cracks.

To this aim, let us consider the following functional dependence for the phenomenon of fatigue crack growth, where renormalized fractal quantities [2] are used instead of the nominal ones:

$$\frac{da^*}{dN} = \Phi\left(\Delta K^*, E, K_{IC}^*, \sigma_y^*, 1-R, h\right) \tag{5}$$

where $\sigma_y^* = \sigma_y h^{1-D}$ is the renormalized yield strength [$FL^{-(D-1)}$], $K_{IC}^* = K_{IC} h^{-\frac{D-1}{2}}$ is the renormalized fracture toughness [$FL^{-(D+2)/2}$], $\Delta K^* = \Delta K a^{-\frac{D-1}{2}}$ is the renormalized stress-intensity factor range [$FL^{-(D+2)/2}$], $a^* = a^D$ is the renormalized crack length [$L^D$], $h$ is a characteristic length [L] (it can be the characteristic specimen size or another parameter with L-dimension). Following the pioneering work [18] and special consideration of this problem in [24], an L-dimension parameter has been included in the analysis. This parameter will be requested for dimension analysis considerations. Euclidean smooth crack profiles have $D=1$ and therefore the previous quantities assume the classical physical dimensions typical of linear elastic fracture mechanics (LEFM).

Considering the Buckingham's $\Pi$ Theorem [18,24] and choosing $\Delta K^*$ and $E$ as the quantities with independent physical dimensions, the number of variables involved in Eq. (5) can be reduced as follows:

$$\frac{da^*}{dN} = \left(\frac{\Delta K^*}{E}\right)^{\frac{2D}{2-D}} \Phi_1\left(K_{IC}^{**}, \sigma_y^{**}, 1-R\right), \tag{6}$$

where $K_{IC}^{**} = \frac{K_{IC}^*}{E h^{\frac{2-D}{2}}}$, $\sigma_y^{**} = \frac{\sigma_y^*}{E h^{1-D}}$ are dimensionless numbers. To derive the numbers $K_{IC}^{**}, \sigma_y^{**}$ we assumed to use the same fractal dimension $D$ for the fracture surface (related to $K_{IC}^{**}$) and for the damage accumulation process (related to $\sigma_y^{**}$). It is not a crucial assumption for the following considerations, but we cannot exclude the necessity of using different fractal dimensions in the equations for $K_{IC}^{**}$, $\sigma_y^{**}$. It is important to note that the dependency of the function $\Phi$ in Eq. (5) on the applied stress range is included in the variable $\Delta K^*$, which defines the oscillatory stress state at the crack tip. As a result of the application of dimensional analysis, the dimensionless function $\Phi_1$ in Eq. (6) becomes independent of $\Delta K^*$ and of the stress range. Moreover, the exponent of $\Delta K^*/E$ in Eq. (6) necessary to have the physical dimensions of the right hand side equal to those of the left hand side is now explicitly dependent on the fractal dimension $D$ of the crack profile, instead of being equal to 2 in case of complete self-similarity, or undetermined in case of incomplete self-similarity (see the classical dimensionless formulation in Eq. (4)).

The relations relating the Euclidean quantities to the renormalized ones

$$a^* = a^D \tag{7a}$$

$$\Delta K^* = \Delta K a^{\frac{D-1}{2}} \tag{7b}$$

can be introduced in Eq. (6) obtaining a crack growth law involving nominal quantities that can be finally compared with the classical Paris' law:

$$D a^{D-1} \frac{da}{dN} = a^{-\frac{(D-1)D}{2-D}} \left(\frac{\Delta K}{E}\right)^{\frac{2D}{2-D}} \Phi_1\left(K_{IC}^{**}, \sigma_y^{**}, 1-R\right) \tag{8}$$

where the chain rule has been applied to differentiate $a^*$ w.r.t. $N$, i.e.:

$$\frac{da^*}{dN} = \frac{da^*}{da}\frac{da}{dN} = D a^{D-1} \frac{da}{dN}.$$

If we call $m$ the exponent of the stress-intensity factor range, then we have the following relation between $m$ and $D$:

$$m = \frac{2D}{2-D}, \quad 1 < D < 2 \tag{9}$$

so that Eq. (8) assumes the following final expression:

$$\frac{da}{dN} = \frac{1}{D} a^{-(D-1)\left(1+\frac{m}{2}\right)} \left(\frac{\Delta K}{E}\right)^m \Phi_1\left(K_{IC}^{**}, \sigma_y^{**}, 1-R\right) = \frac{1}{D} a^{\frac{2(D-1)}{(D-2)}} \left(\frac{\Delta K}{E}\right)^m \Phi_1\left(K_{IC}^{**}, \sigma_y^{**}, 1-R\right) \tag{10}$$

Based on the hypothesis of *incomplete self-similarity* in the dimensionless quantity $1-R$, supported by experimental results discussed in [25, 26], we finally have:

$$\frac{da}{dN} = \frac{1}{D} a^{\frac{2(D-1)}{(D-2)}} \left(\frac{\Delta K}{E}\right)^m (1-R)^\alpha \Phi_1\left(K_{IC}^{**}, \sigma_y^{**}\right) \tag{11}$$

which can be regarded as a generalized Paris' law where the dependency on the crack length is suitably evidenced. It is important to remark here that the crack-size dependency stems directly from the fractality of the rough profiles, without the need of artificially including the crack length among the main parameters affecting fatigue crack growth and then assuming an incomplete similarity in the corresponding dimensionless number. We also note that the exponent $m$ of the Paris law is now related to the morphological properties of the crack profile through Eq. (9). Therefore, we can now state that the incomplete similarity in the crack length (see Eq. (4)) can be explained by the fractality of rough crack profiles. As a result, the incomplete similarity exponent $\alpha_2$ in (4) has now been theoretically related to the fractal dimension D: $\alpha_2 = -(D-1)\left(1+\frac{m}{2}\right) = \frac{2(D-1)}{(D-2)}$. This relation, in addition to the one-to-one correspondence between $m$ and $D$ in Eq. (9), explain the low dependency of the exponents $\alpha_2$ and $m$ on the size of the tested specimen, since the fractal dimension is a size-independent parameter.

It has to be pointed out that the proposed correlation between $m$ and $D$ and the fatigue crack growth equation (11) are valid for LEFM and fractal cracks characterized by a single fractal dimension. Hence, it is necessary to have a structure function with a well-defined power-law trend over at least three orders of magnitude of variation in the sampling interval. Significant deviations from this single power-law regime can lead to erroneous crack growth rate predictions. This could be the case of cracks with bi- or multi-fractal properties, where a single fractal dimension is not enough to describe the structure function of the rough crack profile.

## 3. Integration of the generalized fatigue crack growth law and estimation of the fatigue life

In order to assess the effect of fractality of the rough crack profiles on the fatigue life of an engineering component, it is possible to integrate the generalized Paris' law in Eq. (11) with the crack length ranging from an initial defect size $a_0$ to a critical (final) crack length $a_f$, which corresponds to the failure of the component. Correspondingly, the number of cycles ranges from zero to $N$, the fatigue life of the specimen. In doing that, we focus on a Griffith crack, i.e., a crack of length $2a$ embedded into an infinite plate subjected to cyclic tension. The stress-intensity factor range for self-similar fractal crack range can be related to the applied stress-range as $\Delta K = \chi(D) \Delta \sigma \sqrt{\pi \, a^{2-D}}$, where $\chi(D) = \frac{1}{\pi^{2-D}} \int_0^1 \frac{(1+s)^{2-D} + (1-s)^{2-D}}{(1-s^2)^{(2-D)/2}} \, \mathrm{d}s$ [26]. Other problems can be studied in analogy with smooth cracks ($D=1$), i.e., by including a multiplicative coefficient $k$ dependent on the geometry and loading conditions that can be found in selected handbooks on stress-intensity factors.

Considering the LEFM relationship between $\Delta K$ and $\Delta \sigma$, Eq. (11) can be modified as follows:

$$\frac{\mathrm{d}a}{\mathrm{d}N} = \frac{(1-R)^\alpha}{D} a^{\frac{2(D-1)}{D-2} + \frac{m(2-D)}{2}} \chi(D)^m \Delta\sigma^m \Phi_2\left(K_{IC}^*, \sigma_y^*\right). \tag{12}$$

Equation (12) is an ODE that can be integrated by separating the variables $a$ and $N$. Recalling that $m = 2D/(2-D)$:

$$\frac{\mathrm{d}a}{\mathrm{d}N} = a^{\frac{D^2-2}{D-2}} \frac{(1-R)^\alpha}{D} \chi(D)^m \Delta\sigma^m \Phi_2\left(K_{IC}^*, \sigma_y^*\right).$$

The integration leads to the following result:

$$N = \frac{D}{(1-R)^\alpha \Phi_2\left(K_{IC}^*, \sigma_y^*\right)} \chi(D)^{-m} \Delta\sigma^{-m} \ln\left(\frac{L}{a_0}\right) \quad \text{for} \quad D=1, \tag{13a}$$

$$N = \frac{2-D}{(1-R)^\alpha \Phi_2\left(K_{IC}^*, \sigma_y^*\right) D(D-1)} \chi(D)^{-m} \Delta\sigma^{-m} a_0^{\frac{D(D-1)}{D-2}} \quad \text{for} \quad D>1, \tag{13b}$$

which indicates a logarithmic dependency on the ratio between the structural size, $L$, and the initial crack length, $a_0$, in "classical case" of smooth cracks with dimension $D=1$ and a power law dependency of the type $N \sim a_0^{\frac{D(D-1)}{D-2}}$ for self-similar cracks with fractal dimension $D>1$. Moreover, it also suggests a dependency between the fatigue life and the fractal dimension $D$, for the same values of the applied cyclic load $\Delta\sigma^{-m}$ and of the initial defect size $a_0$.

## 4. Experimental assessment of the proposed fatigue law and comparison with existing models

The fatigue crack growth law (11) suggests that $m$ is an increasing function of $D$. For instance, since $D$ ranges from 1 to 2 for a rough profile, we expect to have $m=2$ and $m=\infty$ as limit cases. This is in good agreement with well-known experimental results, showing a range of $m$ between 2 (ductile materials) and 40 (brittle materials) [20, 22, 28]. Hence, roughness is a decisive property that can be connected to the size of the characteristic material heterogeneity. For quasi-brittle materials, like concrete, the size of the aggregates is much larger than the size of the grain of a polycrystalline metal. Consequently, the crack path is much rougher and the crack growth rate is faster. Unfortunately, no SEM images of fatigue crack paths in brittle or very brittle materials have been found in the literature for performing a quantitative analysis. We also note that Eq. (11) is fully consistent with the fractal scaling law derived in [29, 30] and based on the seminal work in [31].

To provide an experimental assessment, we need to compute the value of *m* from fatigue crack growth data and the value of *D* from the morphology of the corresponding crack paths. The analysis of published experimental results shows a general lack of reports presenting the fractal dimension of the crack surface and the value of *m* simultaneously. Only few papers allow us to calculate these parameters from the published results. For instance, this can be performed using the crack growth data by Hanlon [15] for the Ti specimens cyclically loaded with *R*=0.3 and reported in Fig. 2. The slopes of the Paris' curves are equal to 3.67 and 6.18 for ultrafine grain and coarse grain Ti, respectively. For the same materials we have also analyzed the crack profiles shown in Fig. 1 that provide *D*=1.14 and *D*=1.24 for the ultrafine grain and the coarse grain Ti, respectively. An additional experimental point for Ti-Al-Mn alloy has been obtained by the second author from experimental tests conducted in his laboratory. An indirect estimation of *D* for Aluminum and Steel alloys can be obtained by matching the exponent of the crack size in Eq. (11) with those introduced in the crack growth correlation proposed in [35-37] to capture experimentally evidenced crack-size effects. Although the generalized crack-growth relation in [35-37] was derived under large-scale yielding assumptions, another common explanation of crack-size effects is due to roughness of the crack profiles. Hence, the proposed model based on fractal cracks could also provide an independent interpretation of the anomalous crack-size effects in the short crack regime due to roughness. All the collected values are reported in Tab. 1.

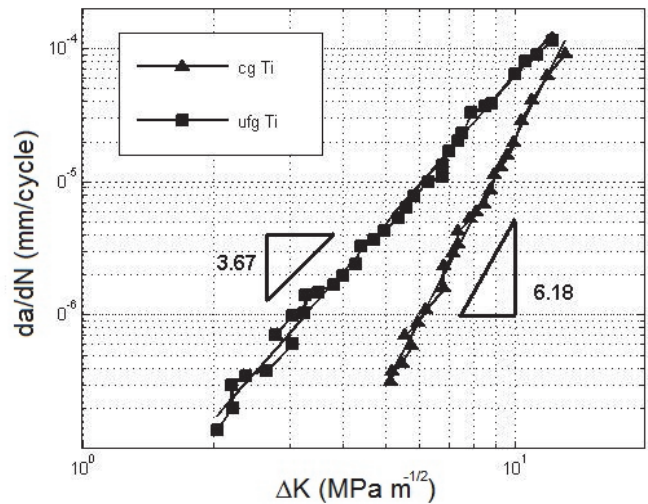

Fig. 2: Paris' curves of coarse grain and ultrafine grain Ti (*R*=0.3), adapted from [12].

| Material | *m* | *D* |
|---|---|---|
| D6ac (Steel) | 2.60 | 1.13 |
| Ti-6Al-4V (Titanium) | 2.50 | 1.11 |
| 7050-T7541 (Aluminum) | 3.00 | 1.20 |
| Micro-alloyed grade C wheel steel | 3.00 | 1.20 |
| 2024-T351 (Aluminum) | 2.84 | 1.17 |

| 2024-T3 (Aluminum) | 3.00 | 1.20 |

Tab. 1: fractal dimensions of fatigue specimens of selected materials estimated from the data in [30-32].

Figure 3 shows the comparison between all the collected values of *m* vs. *D* and the predicted theoretical trend according to Eq. (9) (solid line). The experimental values related to the images by Hanlon [15] are slightly above the theoretical curve, although the trend is correctly reproduced. A possible reason for this discrepancy may arise from the resolution of the digitalized profiles used to compute *D*. In the present study, each profile is a collection of at least 1024 points, thus enabling us to evaluate the structure function in correspondence of nine different sampling intervals, each one half of the next one. More accurate predictions of *D* can be obtained by using higher resolution profile data, permitting to include more length scales into the analysis.

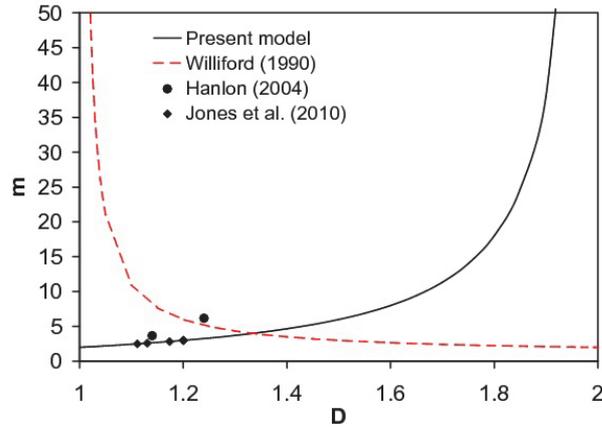

Fig. 3: Exponent *m* vs. fractal dimension *D* (experimental values are reported with filled dots).

As a general trend, the higher the fractal dimension, the higher the slope of the Paris' curve, see the solid curve in Fig. 3. This increasing trend is also qualitatively confirmed by the experimental evidence of fatigue crack growth in concrete. In this material, *D* is usually larger than in metals due to the larger size of the material heterogeneities (the aggregates in this case), with values of about 1.5. The exponent *m* is also much larger than in metals, with values ranging from 5 up to 40 [32], i.e., with a very fast crack growth rate.

Further experimental results are indeed necessary to corroborate the proposed model. However, as it was mentioned before, only in a few studies quantitative fractographic analyses are proposed as, e.g., in [33], where the fractal dimension of crack profiles induced by bending-torsion fatigue in metals has been computed. In that case, however, the fatigue crack growth curves were not determined and the fractal dimension was varying in a very narrow range, from 1.01 to 1.08. The present investigation on the relation between *m* and *D* is therefore the first attempt to correlate the crack morphology to the kinetics of crack propagation.

In Fig. 3 we compare our model predictions with those by the pioneering fractal approach to fatigue by Williford [13] (dashed line). He proposed the following crack-size dependent Paris' law:

$$\frac{da}{dN} = c_1 a^{1-D} \left( \frac{\Delta K}{c_2 E} \right)^{\frac{D}{D-1}} \tag{14}$$

where $c_1$ and $c_2$ are free parameters. The scaling law (14) implies $m = D/(D-1)$ for rough profiles. As a consequence, $m$ should be a decreasing function of $D$ with a vertical asymptote for $D=1$ and the value $m=2$ for $D=2$, see the dashed curve in Fig. 3. This trend is clearly not confirmed by the experimental points and by the values of $m$ and $D$ expected for concrete.

Another interesting feature of Eq. (11) is represented by the crack-size dependency of the multiplying coefficient of $\Delta K$, i.e., the coefficient $C$ of the classical Paris' law. This dependency is also expected according to Eq. (14), but with a different power-law exponent, see Fig. 4. In both models, the exponents of $a$ are negative valued and start from zero for $D=1$, which implies no crack-size effects in case of Euclidean crack profiles. In the present approach, the exponent decreases by increasing $D$ much faster than in the model by Williford. For $D=1.6$, which can be considered as the maximum value of fractal dimension achievable in concrete, the exponent of the crack length in Eq. (11) is 5 times smaller than the value provided by Eq. (14).

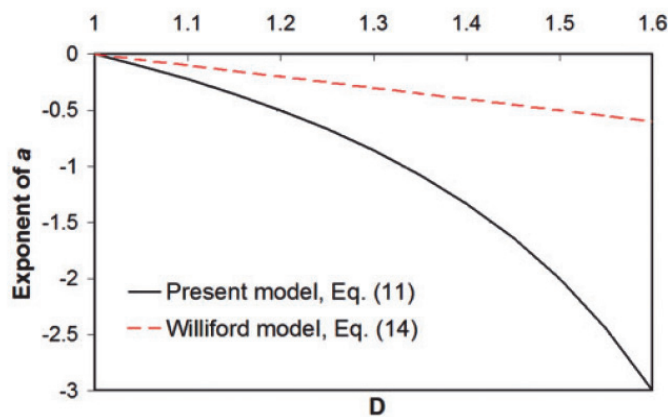

Figure 4. Exponent of the crack length vs. fractal dimension.

Hence, brittle materials like ceramics and concrete with large $D$ are expected to be much more sensible to the initial defect size than metals with $D$ closer to unity. Since the exponent of the crack length is negative valued and $a$ is a quantity much smaller than unity in the short crack regime, the coefficient $C$ computed from Eq. (11) is expected to be an increasing function of the fractal dimension $D$. Moreover, Eq. (14) suggests that short cracks would grow faster than their longer counterparts. An analogous interpretation of crack-size effects in metals is more complex to be made due to crack closure effects induced by large-scale yielding [34-37], a mechanism not present in quasi-brittle materials. Since the present formulation is valid in the LEFM regime, it is not possible to conclude about the role of roughness on crack-size effects. Moreover, roughness of crack profile in ductile materials can be considered as a final result of interaction between plastic deformation zone at crack tip and initial material heterogeneity (inclusions, grain boundaries and so on). In this case the roughness parameters in the present model should be regarded as an overall approximation of the complex processes occurring at the crack tip.

## 5. Conclusion

The theoretical study proposed in the present paper, where dimensional analysis is combined with fractal geometry for the first time, permits to determine a generalized Paris' law with an explicit dependency on the crack length in the framework of LEFM. The derived scaling law is in agreement with that proposed by the first author in previous studies according to pure fractal geometry considerations. As a main outcome, the self-similarity exponent of the crack length and the exponent $m$ of the Paris' law are theoretically related to the fractal dimension of the fatigue crack profiles. Since the fractal dimension is a scale-invariant parameter, this explains the low dependency of these exponents on the size of the tested samples. Previous work on this matter by Williford [13] suggested a different relation between *m* and *D*. A preliminary comparison with experimental results shows that our proposed model is able to capture the experimental trend, which is however not captured by the fractal model by Williford. In particular, the exponent *m* is found to be an increasing function of the fractal dimension *D*. This is also consistent with the very fast crack growth rate in concrete and ceramics (*m* ranging from 5 to 40) whose crack profiles have in general higher fractal dimensions than those of metals due to the presence of bigger heterogeneities and grain sizes.

A complete analysis of crack-size effects in metals is left for further investigation. Plasticity effects, which have not been included in the present model, are indeed important and can influence the obtained trends. From dimensional analysis considerations, we found that the ratio $Z \sim \sqrt{\frac{h}{r_y}}$, where $r_y = K_{IC}^2 / \sigma_y^2$, enters the expression of the dimensionless function $\Phi_1\left(K_{Ic}^{**}, \sigma_y^{**}, 1-R\right)$. The number $Z$ influences the fracture mechanism at the crack tip: small values of $Z$ (less than 2, according to [24]) correspond to large-scale yielding. On the other hand, large values of $Z$ are associated with high strength/low toughness materials where LEFM conditions prevail. Therefore, an effect of plasticity on the Paris' law parameter *C* is envisaged. Plasticity-induced crack closure effects are also affecting the Paris' law exponent *m* and their contribution has to be suitably accounted for by considering large-scale yielding as in [34].

Integrating the crack-size dependent fatigue crack growth equation, an estimation of the fatigue life of a structural component has also been provided. We found that a logarithmic dependency of the cycles to failure on the ratio between the specimen size and the length of pre-existing micro defects can be observed for smooth cracks only (with fractal dimension $D = 1$). In general (for self-similar cracks with dimension $D > 1$) we expect a power law dependency of the type $N \sim a_0^{\frac{D(D-1)}{D-2}}$. Moreover, the fatigue life is proportional to the fractal dimension *D*, provided that the applied cyclic load $\Delta \sigma^{-m}$ is kept constant from a specimen to another during the tests.

It has to be remarked that the proposed relation (11) is applicable to fatigue crack morphologies characterized by a single fractal dimension *D*. In case of modifications of the crack morphology during crack propagation, as it may occur in case of overloading effects, the values of *D* entering Eq. (11) should be consistently updated during a crack growth simulation to obtain accurate results.

Starting from these promising results, more detailed investigations using higher resolution profiles is indeed required. This would also permit to assess the range of validity of fractality, as well as the appearance of multifractal regimes characterized by different fractal dimensions. In particular, the upper and lower cut-off lengths to the fractal power-law regime are important to be quantified. From one side, fractality may be violated at the nanoscale, where also continuum mechanics breaks down and crack propagation becomes a discrete phenomenon of atoms separation. On the opposite side, the size-scale of the sample may also provide a cut-off to the power-law regime, with the appearance of important size-scale effects.

**Acknowledgements**

The research leading to these results has received funding from the European Research Council under the European Union's Seventh Framework Programme (FP/2007–2013)/ERC Grant Agreement No. 306622 (ERC Starting Grant ''Multi-field and multi-scale Computational Approach to Design and Durability of PhotoVoltaic Modules'' – CA2PVM). The support of the Italian Ministry of Education, University and Research to the Project FIRB 2010 Future in Research ''Structural mechanics models for renewable energy applications'' (RBFR107AKG) is also gratefully acknowledged. Dr. Plekhov acknowledges the partial financial support of grant МД-2684.2012.1 during this work.